\documentstyle[prl,aps,epsfig,floats]{revtex}
\begin{document}
\draft
\title{Chaotic Transport and Current Reversal in Deterministic Ratchets} 
\author{Jos\'e L. Mateos}
\address{Instituto de F\'{\i}sica, Universidad Nacional Aut\'onoma de
M\'exico,\\
Apartado Postal 20-364, M\'exico D.F. 01000, M\'exico}

\maketitle

\begin{abstract}
We address the problem of the classical deterministic dynamics of a 
particle in a periodic asymmetric potential of the ratchet type. We 
take into account the inertial term in order to understand the role 
of the chaotic dynamics in the transport properties. By a comparison 
between the bifurcation diagram and the current, we identify the 
origin of the current reversal as a bifurcation from a chaotic to a 
periodic regime. Close to this bifurcation, we observed trajectories
revealing intermittent chaos and anomalous deterministic diffusion.
\end{abstract}

\vspace{0.8cm}

In recent years there has been an increasing interest in the
study of the transport properties of nonlinear systems that can 
extract usable work from unbiased nonequilibrium fluctuations. 
These, so called ratchet systems, can be modeled, for instance,
by considering a Brownian particle in a 
periodic asymmetric potential and acted upon by an external 
time-dependent force of zero average \cite{hanggi96,reviews}.
This recent burst of work is motivated in part by the challenge
to explain the unidirectional transport of molecular motors 
in the biological realm \cite{bios}. Another source of motivation
arises from the potential for new methods of separation
or segregation of particles \cite{segre}, and more recently in 
the recognition of the ``ratchet effect'' in the quantum domain 
\cite{quantum}. The latter research includes: a quantum ratchet 
based on an asymmetric (triangular) quantum dot \cite{qdot};
an asymmetric antidot array \cite{antidot}; the ratchet effect in
surface electromigration \cite{surface}; a ratchet potential
for fluxons in Josephson-junctions arrays \cite{joseph};
ratchet effect in cold atoms using an asymmetric optical lattice
\cite{atoms}; and the reducing of vortex density in superconductors
using the ratchet effect \cite{vortex}. 

In order to understand the generation of unidirectional motion
from nonequilibrium fluctuations, several models have been used. 
In Ref. \cite{hanggi96}, there is a classification of different 
types of ratchet systems; among them we can mention the 
``Rocking Ratchets'', in which the particles move in an asymmetric 
periodic potential subject to spatially uniform, time-periodic 
deterministic forces of zero average.
Most of the models, so far, deal with the overdamped case in 
which the inertial term due to the finite mass of the particle is 
neglected. However, in recent studies, this oversimplification was 
overcome by treating properly the effect of finite mass 
\cite{quantum,jung96,inertial}. 

In particular, in a recent paper \cite{jung96}, Jung, 
Kissner and H\"{a}nggi study the effect of finite inertia 
in a deterministically rocked, periodic ratchet potential.
They consider the deterministic case in which noise is 
absent \cite{dete}. The inertial term allows the possibility of 
having both regular and chaotic dynamics, and this deterministically 
induced chaos can mimic the role of noise. They showed that the 
system can exhibit a current flow in either direction, presenting 
multiple current reversals as the amplitude of the external force 
is varied. 

In this paper, the problem of transport in periodic asymmetric 
potentials of the ratchet type is address. We elaborate on 
the model analyzed by Jung {\it et al.} \cite{jung96}, 
in which they find multiple current reversals in the dynamics. 
In fact, the study of the current-reversal phenomena has given
rise to a research activity on its own \cite{hanggi96}.
 
The goal of this paper is to reveal the origin of the current 
reversal, by analyzing in detail the dynamics for 
values of the parameters just before and after the critical 
values at which the current reversal takes place.

Let us consider the one-dimensional problem of a particle driven by a
periodic time-dependent external force, under the influence of an 
asymmetric periodic potential of the ratchet type. The time average 
of the external force is zero. Here, we do not take into account 
any kind of noise, and thus the dynamics is deterministic. 
The equation of motion is given by

\begin{equation}
m\ddot x + \gamma \dot x + {\frac{dV(x)}{dx}} = F_0 \cos(\omega_D t),
\end{equation}

\noindent where $m$ is the mass of the particle, 
$\gamma$ is the friction coefficient,
$V(x)$ is the external asymmetric periodic potential, $F_0$ is the amplitude
of the external force and $\omega_D$ is the frequency of the external
driving force. The ratchet potential is given by 

\begin{equation}
V(x) = V_1 - V_0 \sin {\frac{2\pi (x-x_0)}{L}} - {\frac{V_0}{4}}
\sin {\frac{4\pi (x-x_0)}{L}},
\end{equation}

\noindent where $L$ is the periodicity of the potential, 
$V_0 $ is the amplitude, and $V_1$ is an arbitrary constant. The potential
is shifted by an amount $x_0$ in order that the minimum of the potential
is located at the origin.

Let us define the following dimensionless units: ${x^{\prime }=x/L}$,
$x_{0}^{\prime }=x_{0}/L$, $t^{\prime }=\omega _{0}t$,
$w=\omega _{D}/\omega_{0}$, $b=\gamma /m\omega _{0}$ and
$a=F_{0}/mL\omega _{0}^{2}$. 
Here, the frequency $\omega _{0}$ is given by $\omega _{0}^{2}=4\pi
^{2}V_{0}\delta /mL^{2}$ and $\delta $ is defined by $\delta =\sin (2\pi
|x_{0}^{\prime }|)+\sin (4\pi |x_{0}^{\prime }|)$. 

\begin{figure}[htb]
\centerline{\epsfig{file=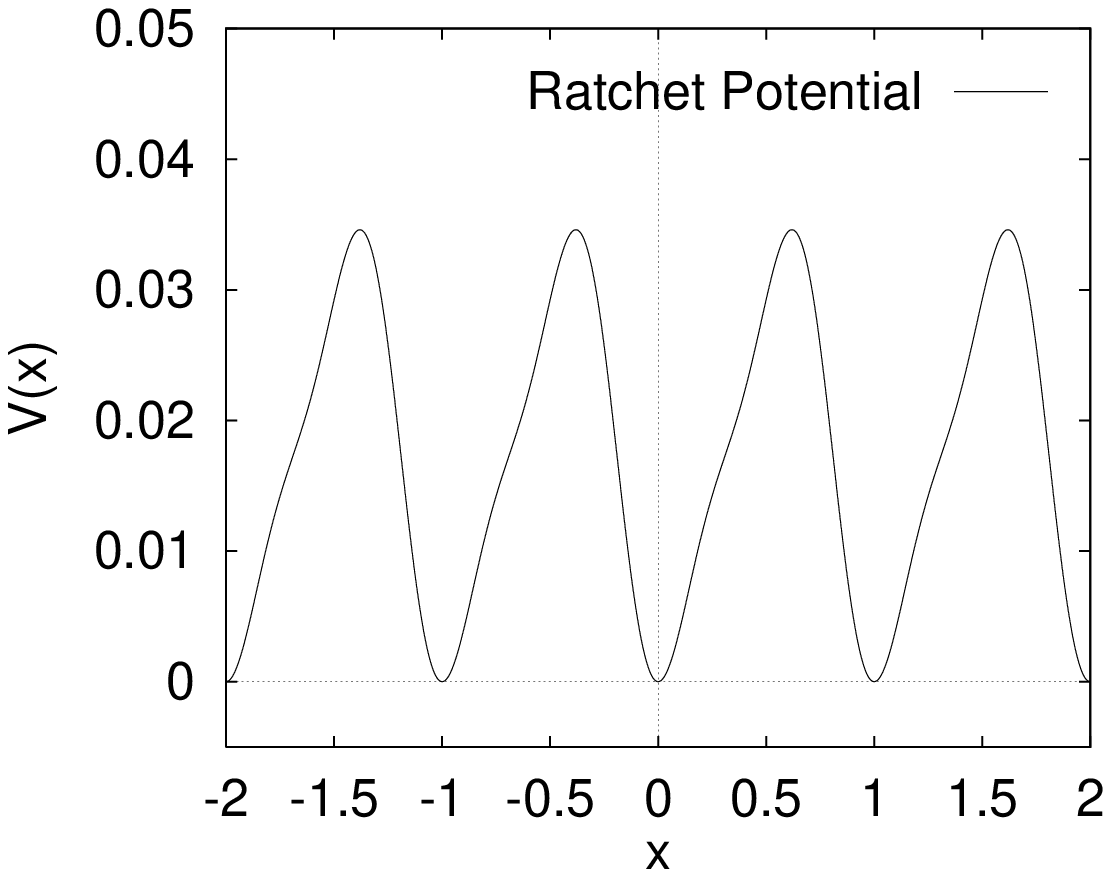,width=7.0cm}}
\caption{The dimensionless ratchet periodic potential $V(x)$.}
\label{fig1}
\end{figure}

The frequency $\omega _{0}$ is the frequency of the linearize motion around
the minima of the potential, thus we are scaling the time with the natural
period of motion $\tau _{0}=2\pi /\omega _{0}$. The dimensionless equation
of motion, after renaming the variables again without the primes, becomes 

\begin{equation}
\ddot{x}+b\dot{x}+{\frac{dV(x)}{dx}}=a\cos (wt),    \label{eqmov}
\end{equation}

\noindent where the dimensionless potential \cite{const} is given by
${V(x)=C-(\sin 2\pi (x-x_{0}) + 
0.25 \sin 4\pi (x-x_{0}))/4\pi^{2}\delta}$ 
and is depicted in Fig.\ \ref{fig1}.

In the equation of motion Eq. (\ref{eqmov}) there are three
dimensionless parameters: $a$, $b$ and $w$, defined above in terms of
physical quantities. We vary the parameter $a$ and fix 
$b = 0.1$ and $w = 0.67$ throughout this paper.

The extended phase space in which the dynamics is taking place is
three-dimensional, since we are dealing with an inhomogeneous
differential equation with an explicit time dependence. This equation
can be written as a three-dimensional dynamical system, that we solve 
numerically, using the fourth-order Runge-Kutta algorithm.
The equation of motion Eq. (\ref{eqmov}) is nonlinear and thus allows the
possibility of chaotic orbits. If the inertial term associated 
with the second derivative $\ddot{x}$ were absent, then the 
dynamical system could not be chaotic.

The main motivation behind this work is to study in detail the origin of
the current reversal in a chaotically deterministic rocked ratchet.
In order to do so, we have to study first the current $J$ itself, 
that we define as the time average of the average velocity over an 
ensemble of initial conditions. Therefore,
the current involves two different averages: the first average is
over $M$ initial conditions, that we take equally distributed in
space, centered around the origin and with an initial velocity
equal to zero. For a fixed time, say $t_j$, we obtain an average
velocity, that we denoted as $v_j$, and is given by $ v_j = 
{\frac{1}{M}}\sum\limits_{i=1}^M {\dot{x_i}}(t_j)$.
The second average is a time average; since we take a discrete time
for the numerical solution of the equation of motion, we have a
discrete finite set of $N$ different times $t_j$; 
then the current can be defined as $ J = 
{\frac{1}{N}}\sum\limits_{j=1}^N {v_j}$.
This quantity is a single number for a fixed set of parameters 
${a,b,w}$, but it varies with the parameter $a$, fixing $b$ and $w$.

Besides the continuum orbits in the extended phase space, 
we can obtain the Poincar\'e section,
using as a stroboscopic time the period of oscillation of the
external force. With the aid of Poincar\'e sections we can
distinguish between periodic and chaotic orbits, and we can obtain
a bifurcation diagram as a function of the parameter $a$.

The bifurcation diagram for $b=0.1$ and $w=0.67$ is shown in
Fig. 2a in a limited range of the parameter $a$. We can
observe a period-doubling route to chaos and after a
chaotic region, there is a bifurcation taking place at
a critical value $a_c\simeq 0.08092844$. It is precisely at 
this bifurcation point that the current reversal occurs. 
After this bifurcation, a periodic window emerges, with an 
orbit of period four. In Fig. 2b, we show the current as a 
function of the parameter $a$, in exactly the same range as 
the bifurcation diagram above. We notice the abrupt transition 
at the bifurcation point that leads to the first current 
reversal. In Figs. 2a,b we are analyzing only a short range of
values of $a$, where the first current reversal takes place.
If we vary $a$ further, we can obtain multiple current 
reversals \cite{jung96}.

In order to understand in more detail the nature of the current reversal, 
let us look at the orbits just before and after the transition. The 
reversal occurs at the critical value $a_c\simeq  0.08092844$. 
If $a$ is below this critical value $a_c$, say $a = 0.074$, 
then the orbit is periodic, with 
period two. For this case we depict, in Fig. 3a, the
position of the particle as a function of time. We notice a period-two 
orbit, as can be distinguish in the bifurcation diagram for $a = 0.074$. 
In Fig. 3b we show again the position as a function of time for 
$a = 0.081$, which is just above the critical value $a_c$. In this case, 
we observe a period-four orbit, that corresponds to the periodic
window in the bifurcation diagram in Fig. 2a. This orbit is
such that the particle is ``climbing'' in the negative direction, 
that is, in the direction in which the slope of the potential is higher.
We notice that there is a qualitative difference
between the periodic orbit that transport particles to the positive 
direction and the periodic orbit that transport particles to the 
negative direction: in the latter case, the particle requires twice
the time than in the former case, to advances one well in the 
ratchet potential. A closer look at the trajectory in Fig. 3b reveals 
the ``trick'' that the particle uses to navigate in the negative 
direction: in order to advance one step to the left, it moves first 
one step to the right and then two steps to the left. The net result is 
a negative current.

In Fig. \ref{fig4}, we show a typical trajectory for $a$ just below 
$a_c$. The trajectory is chaotic and the corresponding 
chaotic attractor is depicted in Fig. 5. 
In this case, the particle starts at the origin with no velocity; it 
jumps from one well in the ratchet potential to another well to the right 
or to the left in a chaotic way. The particle gets trapped oscillating for 
a while in a minimum (sticking mode), as is indicated by the integer 
values of $x$ in the ordinate, and suddenly starts a running mode with 
average constant velocity in the negative direction. In terms of the 
velocity, these running modes, as the one depicted in Fig. 3b, correspond 
to periodic motion. The phenomenology can be described as follows. 
For values of $a$ above $a_c$, as in 
Fig. 3b, the attractor is a periodic orbit. For $a$ slightly less than
$a_c$ there are long stretches of time (running or laminar modes) during 
which the orbit appears to be periodic and closely resembles the orbit 
for $a>a_c$, but this regular (approximately periodic) behavior is 
intermittently interrupted by finite duration ``bursts'' in which the
orbit behaves in a chaotic manner. The net result in the velocity 
is a set of periodic stretches of time interrupted by burst of 
chaotic motion, signaling precisely the phenomenon of intermittency 
\cite{inter}. As $a$ approach $a_c$ from below, the duration of the
running modes in the negative direction increases, until the duration
diverges at $a=a_c$, where the trajectory becomes truly periodic.

To complete this picture, in Fig. \ref{fig5}, we show two attractors: 
1) the chaotic attractor for $a=0.08092$, just below $a_c$, 
corresponding to the trajectory in Fig. \ref{fig4} and; 
2) the period-4 attractor for $a=0.08093$, corresponding to the 
trajectory in Fig. 3b. This periodic attractor consist of four 
points in phase space, which are located at the center of
the open circles. We obtain these attractors confining the dynamics 
in $x$ between $-0.5$ and $0.5$. As $a$ approaches $a_c$ from below, 
the dynamics in the attractor becomes intermittent, spending most of the
time in the vicinity of the period-4 attractor, and suddenly ``jumping''
in a chaotic way for some time, and then returning close to the period-4
attractor again, and so on. In terms of the velocity, the result is an
intermittent time series as discussed above.

In order to characterize the deterministic diffusion in this regime, 
we calculate the mean square displacement $\langle x^2 \rangle$ 
as a function of time. We obtain numerically that 
$\langle x^2 \rangle \sim t^{\alpha}$, where the
exponent $\alpha \simeq 3/2$. This is a signature of anomalous
deterministic diffusion, in which $\langle x^2 \rangle$ grows faster
than linear, that is, $\alpha > 1$ (superdiffusion). Normal 
deterministic diffusion corresponds to $\alpha = 1$.
In contrast, the trajectories in Figs. 3a and 3b transport 
particles in a ballistic way, with $\alpha = 2$.
The relationship between anomalous deterministic diffusion and
intermittent chaos has been explored recently, together with 
the connection with L\'evy flights \cite{levy}. The character
of the trajectories, as the one in Fig. \ref{fig4}, remains to 
be analyzed more carefully in order to determine if they 
correspond to L\'evy flights.

In summary, we have identify the mechanism by which the current 
reversal in deterministic ratchets arises: it corresponds to 
a bifurcation from a chaotic to a periodic regime. Near this 
bifurcation, the chaotic trajectories exhibit intermittent 
dynamics and the transport arises through deterministic 
anomalous diffusion with an exponent greater than one 
(superdiffusion). As the control parameter $a$ approaches the 
critical value $a_c$ at the bifurcation from below, the duration
of the running modes in the negative direction increases. 
Finally, the duration diverges at the critical value, leading
to a truly periodic orbit in the negative direction. This is
precisely the mechanism by which the current-reversal takes 
place.

The author acknowledges helpful discussions with 
P. H\"{a}nggi, P. Jung, G. Cocho, C. Garc\'{\i}a, H. Larralde,
G. Mart\'{\i}nez-Mekler, V. Romero-Roch\'{\i}n and F. Leyvraz.

\vfill\eject

\vfill\eject

\begin{figure}[htb]
\centerline{\epsfig{file=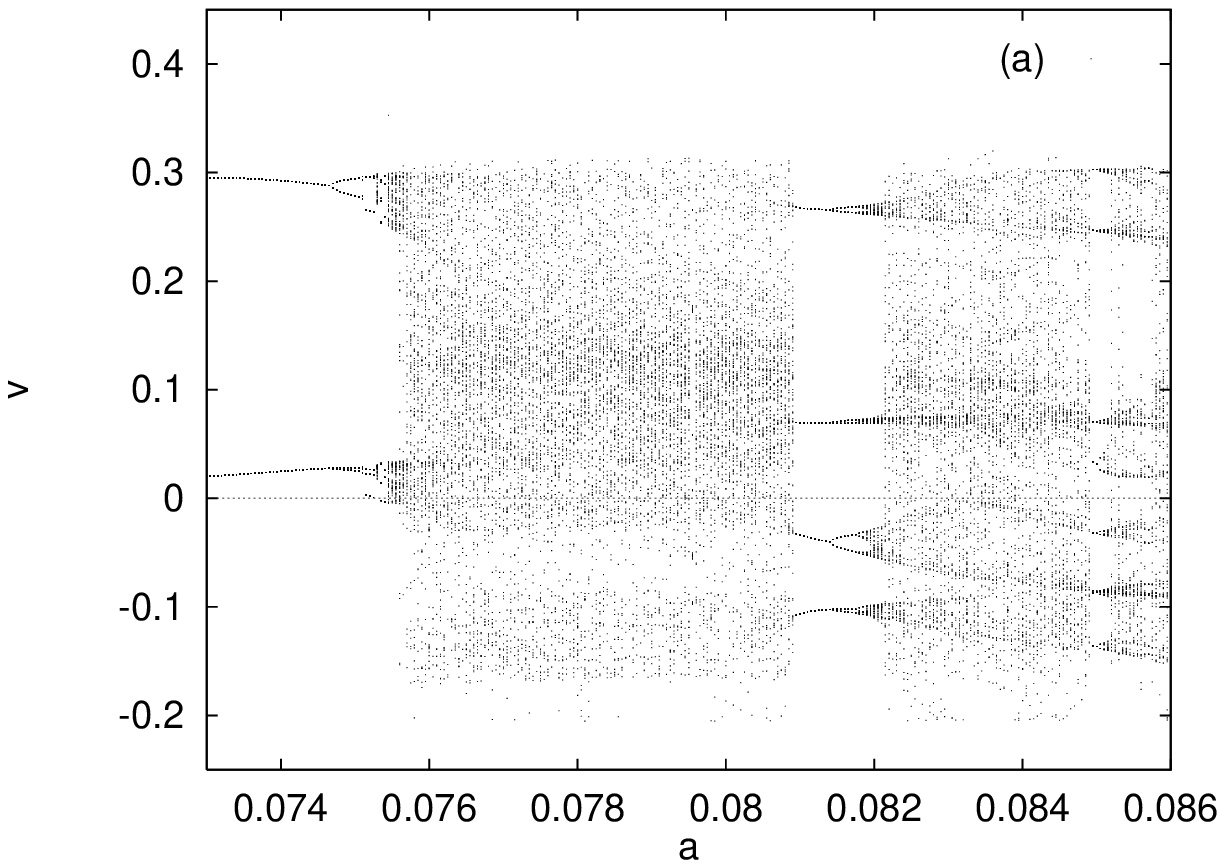,width=8.6cm}}
\label{fig2a}
\end{figure}

\begin{figure}[htb]
\centerline{\epsfig{file=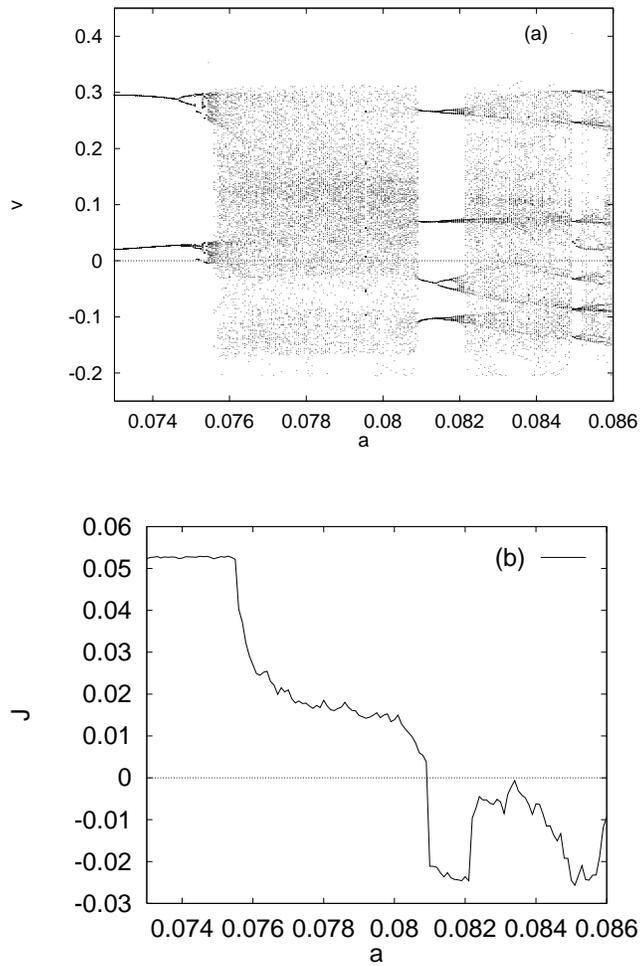,width=8.6cm}}
\caption{For $b=0.1$ and $w=0.67$ we show: (a) The bifurcation
diagram as a function of $a$; (b) The current $J$ as a
function of $a$. The range in the parameter $a$ corresponds to
the first current reversal.}
\label{fig2b}
\end{figure}

\vfill\eject

\begin{figure}[htb]
\centerline{\epsfig{file=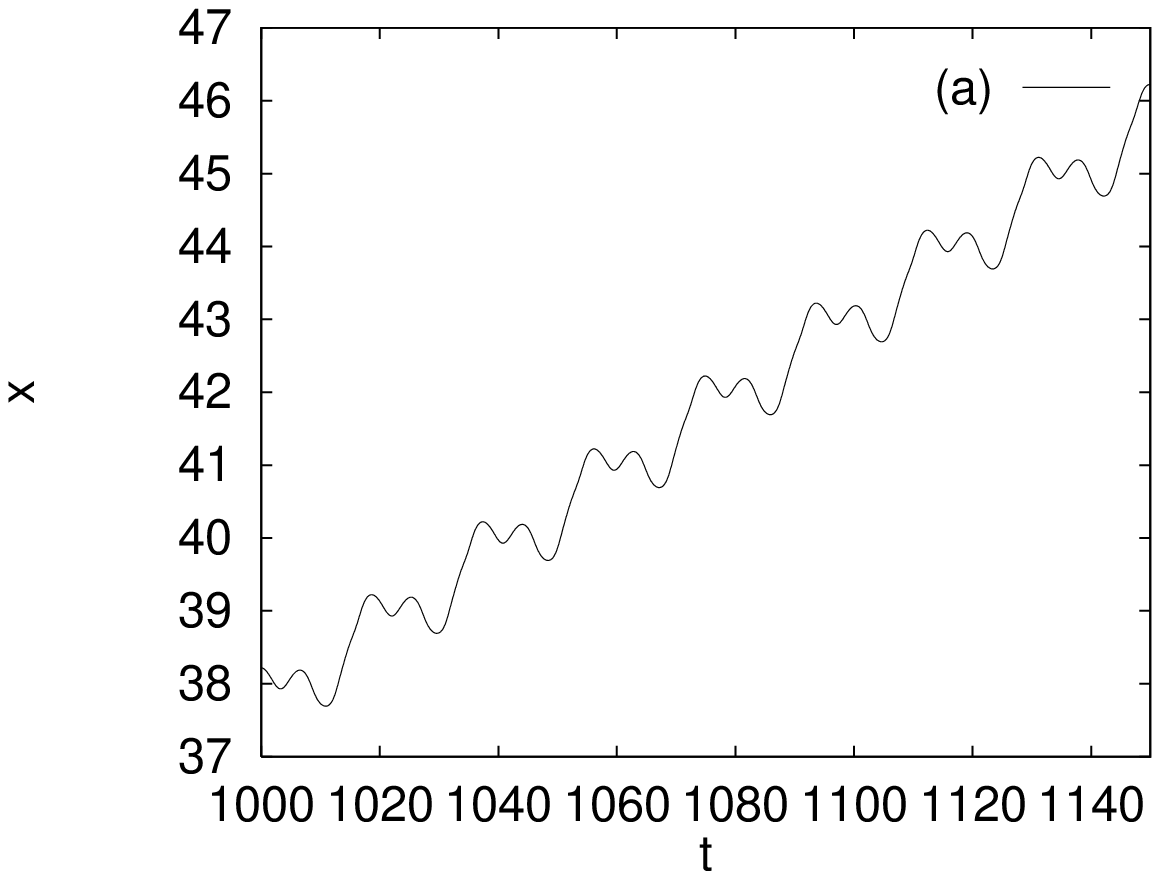,width=7.0cm}}
\label{fig3a}
\end{figure}

\begin{figure}[htb]
\centerline{\epsfig{file=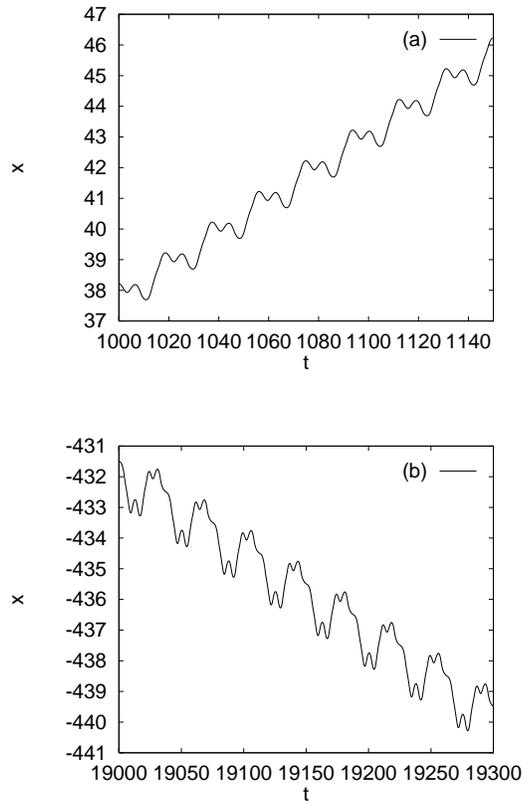,width=7.0cm}}
\caption{For $b=0.1$ and $w=0.67$ we show: 
(a) The trajectory of the particle for $a = 0.074$ (positive current); 
(b) The trajectory for $a = 0.081$ (negative current).}
\label{fig3b}
\end{figure}

\vfill\eject

\begin{figure}[htb]
\centerline{\epsfig{file=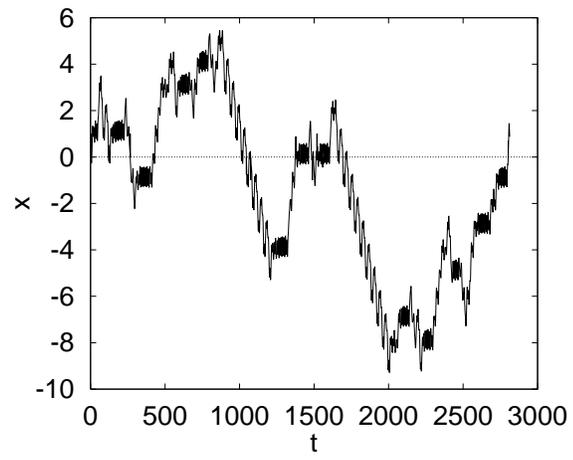,width=8.6cm}}
\caption{The intermittent chaotic trajectory of the particle for 
$b=0.1$, $w=0.67$ and $a = 0.08092844$.}
\label{fig4}
\end{figure}

\vfill\eject

\begin{figure}[htb]
\centerline{\epsfig{file=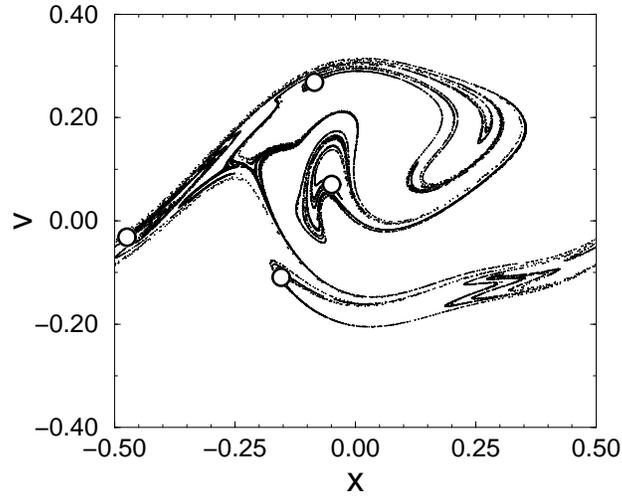,width=8.6cm}}
\caption{For $b=0.1$ and $w=0.67$ we show two attractors: a chaotic
attractor for $a=0.08092$, just below $a_c$, and; a period-4   
attractor consisting of four points located at the center of the
open circles.}
\label{fig5}
\end{figure}

\vfill\eject


\begin{references}
\bibitem{const}
The constant $C$ is such that $V(0)=0$, and is given by 
$C = -(\sin 2\pi x_{0} + 0.25 \sin 4\pi x_{0})/4\pi^{2}\delta$.
In this case, $x_{0}\simeq -0.19$, $\delta \simeq 1.6$ and 
$C\simeq 0.0173$.
\end{references}
\end{document}